\documentstyle[epsf,axodraw]{article} 
\begin{document}

%%%%%%%%%%%%%%%%%%%%%%%%%%%%%Start of Text%%%%%%%%%%%%%%%%%%%%%%%%%%%%%%%%%
\bigskip
{\Large\bf
\centerline{Cancellation of Infrared Divergences in Hadronic} 
\centerline{Annihilation Decays of Heavy Quarkonia}
\bigskip
\normalsize

\centerline{Han-Wen Huang$^{1,2}$~~~~Kuang-Ta Chao$^{1,3}$}
\centerline{\sl $^1$ CCAST (World Laboratory), Beijing 100080, P.R.China} 
\centerline{\sl $^2$ Institute of Theoretical Physics, Academia Sinica,
      P.O.Box 2735,}
\centerline{\sl Beijing 100080, P.R.China}
\centerline{\sl $^3$ Department of Physics, Peking University, Beijing 100871,
           P.R.China}
\bigskip

\begin{abstract}
In the framework of a newly developed factorization formalism which is based 
on NRQCD, explicit cancellations are shown for the infrared divergences that 
appeared in the previously calculated hadronic annihilation decay rates of 
P-wave and D-wave heavy quarkonia. We extend them to a more general case 
that to leading order in $v^2$ and next-to-leading order in $\alpha_s$, the 
infrared divergences in the annihilation amplitudes of color-singlet 
$Q\bar{Q}(^{2S+1}L_J)$ pair can be 
removed by including the contributions of color-octet operators 
$Q\bar{Q}(^{2S+1}(L-1)_{J^{\prime}})$, $Q\bar{Q}(^{2S+1}(L-3)_{J^{\prime\prime}})$, $\cdots$ in NRQCD. We also give the decay widths 
of $^3D_J\rightarrow LH$ at leading order in $\alpha_s$.
\end{abstract}

\vfill\eject\pagestyle{plain}\setcounter{page}{1}

Since the discovery of charmonium and bottonium, heavy quarkonium physics has
drawn much attention from theorists. Particularly, studying  
the decay rates of heavy quarkonium states into light hadrons may provide very 
interesting tests of perturbative quantum chromodynamics (QCD).   
In the early ``color singlet model'' analysis, quarkonium is considered as a bound state of color singlet $Q\bar{Q}$ pair which is in a fixed 
angular-momentum state. And in the formula of the decay rates, all 
long-distance nonperturbative part was assumed to be factored into the nonrelativistic 
wavefunction of color singlet $Q\bar{Q}$ or its derivative at the origin, and 
the perturbative part is related to the annihilation of color-singlet 
$Q\bar{Q}$ which can be calculated using perturbative QCD. In the 
nonrelativistic limit, this early factorization formalism was supported by 
explicit calculations for S-wave decay at next-to-leading order in 
$\alpha_s$ \cite{hag}. But in the cases of P-wave \cite{Barbi} and D-wave 
\cite{dwave} quarkonium decays, the infrared divergences occur, which are
associated with the logarithms of binding energy in the perturbative
calculations of color-singlet $Q\bar{Q}$ annihilation amplitudes.
These indicate that the decay rates are sensitive to nonperturbative effects 
beyond those related to the wavefunction of color-singlet $Q\bar{Q}$ pair or
its derivative at the origin, and not all nonperturbative effects can be 
factored into the color-singlet component of quarkonium. The early factorization
formalism is incomplete and fails to give correct results in some cases.    

Recently, Bodwin, Braaten and Lepage have developed a rigorous factorization
formalism \cite{BBL} which is based on an effective field theory, nonrelativistic QCD (NRQCD). 
This factorization formalism gives a systematic analysis for the decay and
production of heavy quarkonium to any given order in $v^2$ and $\alpha_s$.
In this factorization formalism, the decay rates can be written as a sum of a 
set of long-distance nonperturbative matrix elements, each of which multiplied 
by a short-distance perturbative coefficient, which can be calculated in 
perturbative QCD. The nonperturbative factors in this formalism are defined 
rigorously in NRQCD. This factorization formula applies to the decay of S-wave,
P-wave  equally as well as to the decay of any high orbital angular-momentum 
states. The essential point of this formalism is that quarkonium is treated as 
a state consisting of $Q\bar{Q}$, $Q\bar{Q}g$ and other high Fock components 
rather than only consisting of color-singlet $Q\bar{Q}$ component. In some 
cases the color-octet component even give dominant contribution to the decay.

In NRQCD the annihilation rate of a quarkonium state
$H$ to light hadrons ($LH$) can be written as
\begin{eqnarray}\label{rate}\nonumber
\Gamma(H\rightarrow LH)&=&2Im<H|\delta{\cal L}_{4-fermion}|H>\\
&=&\sum_n\frac{Imf_n(\alpha_s)}{m^{d_n-4}}<H|{\cal O}_n|H>.
\end{eqnarray}
where the sum is over all possible local 4-fermion operators ${\cal O}_n$
that annihilate and create a $Q\bar{Q}$ pair, and $d_n$ is the scaling
dimension of ${\cal O}_n$.

Since all nonperturbative effects are factored into the matrix elements, 
the coefficients must be infrared finite and can be calculated in perturbative 
QCD. The coefficients associated with S-wave decay to next-to-leading 
order both in $\alpha_s$ and $v^2$ have been given in Ref.\cite{BBL,ktchao}. There the calculation
of $Q\bar{Q}$ annihilation rates in full perturbative QCD does not involve 
infrared divergences, and in deriving the color-singlet coefficients one does not
need to consider the contribution from color-octet operators in NRQCD, 
which is suppressed
by $v^4$ as compared with that from dominant color-singlet operator in the S-wave
decay. But for the P-wave and D-wave decays, the color-octet component may contribute at
the same order in $v^2$ as the color-singlet component, and we would get infrared 
divergence coefficients if only color-singlet's contribution is considered in NRQCD.
It is very interesting to know how the cancellation of these divergences occurs  
by introducing the color octet operators in NRQCD and get infrared finite
coefficients. In this paper we will show the explicit cancellation
of infrared divergences by analysing the new factorization formulism to P-wave and D-wave decay processes.
The decay rates of D-wave triplet states will also be given to leading order in
$\alpha_s$.

It is known that the coefficients $Imf_n(\alpha_s)$ in (\ref{rate}) can be 
determinated by matching the imaginary part of $Q\bar{Q}$ forward scattering 
amplitude in the full theory with that in NRQCD, where $Q\bar{Q}$ is in the corresponding 
color and angular momentum state. In the full theory we recalculate the scattering 
amplitudes of color-singlet P-wave and D-wave $Q\bar{Q}$ in dimensional 
regularization scheme and get
\begin{equation}\label{3pf}
Im{\cal M}(^3P_J)_{full~QCD}=\frac{2n_fC_F\alpha_s^3}{9N_cm^6}(-\frac{1}{2\epsilon_{IR}})+C(^3P_J),
\end{equation}
\begin{equation}
Im{\cal M}(^1P_1)_{full~QCD}=\frac{(N_c^2-4)C_F\alpha_s^3}{3N_c^2m^6}
(-\frac{1}{2\epsilon_{IR}})+C(^1P_1),
\end{equation}
\begin{equation}
Im{\cal M}(^3D_1)_{full~QCD}=\frac{76(N^2_c-4)C_F\alpha_s^3}{135N^2_cm^6}
(-\frac{1}{2\epsilon_{IR}})+C(^3D_1),
\end{equation}
\begin{equation}
Im{\cal M}(^3D_2)_{full~QCD}=\frac{(N^2_c-4)C_F\alpha_s^3}{15N^2_cm^6}
(-\frac{1}{2\epsilon_{IR}})+C(^3D_2),
\end{equation}
\begin{equation}\label{3d3f}
Im{\cal M}(^3D_3)_{full~QCD}=\frac{4(N^2_c-4)C_F\alpha_s^3}{15N^2_cm^6}
(-\frac{1}{2\epsilon_{IR}})+C(^3D_3).
\end{equation}
Here for convenience we only give their infrared divergence part and $C(^{2S+1}L_J)$
represent finite terms. We find when 
making a substitution $\frac{-1}{2\epsilon_{IR}}\rightarrow
ln\frac{m}{\varepsilon}$, the above results will coincide with that derived in
\cite{Barbi} and \cite{dwave}. We use dimensional regularization to control
infrared divergences and take quarks to be on-shell in order to keep gauge 
invariance. Because  conventional NRQCD is treated under the on-shell condition, we 
must use the same regularization scheme in the full theory and NRQCD so that we 
can get a compatible result.

\begin{center}\begin{picture}(300,60)(0,0)
\ArrowLine(100,50)(150,30)\ArrowLine(200,10)(150,30)
\ArrowLine(150,30)(100,10)\ArrowLine(150,30)(200,50)
\Vertex(150,30){2}
\end{picture}

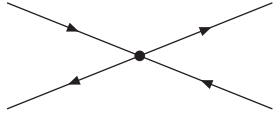
{Fig.1~~Feynman diagram contributing to $Im{\cal M}(^{2S+1}L_J)
_{NRQCD}$ through the color-singlet operator ${\cal O}_1(^{2S+1}L_J)$.}

\end{center}
 
\begin{center}\begin{picture}(300,120)(0,0)
\ArrowLine(10,110)(75,90)\ArrowLine(140,70)(75,90)
\ArrowLine(75,90)(140,110)\ArrowLine(75,90)(10,70)
\Gluon(42.5,100)(107.5,100){1.5}{15}
\Vertex(75,90){2}
\Text(75,65)[]{(a)}

\ArrowLine(160,110)(225,90)\ArrowLine(290,70)(225,90)
\ArrowLine(225,90)(290,110)\ArrowLine(225,90)(160,70)
\Gluon(192.5,80)(257.5,80){1.5}{15}
\Vertex(225,90){2}
\Text(225,65)[]{(b)}

\ArrowLine(10,50)(75,30)\ArrowLine(140,10)(75,30)
\ArrowLine(75,30)(140,50)\ArrowLine(75,30)(10,10)
\GlueArc(75,30)(20,344,164){2}{15}
\Vertex(75,30){2}
\Text(75,5)[]{(c)}

\ArrowLine(160,50)(225,30)\ArrowLine(290,10)(225,30)
\ArrowLine(225,30)(290,50)\ArrowLine(225,30)(160,10)
\GlueArc(225,30)(20,16,196){2}{15}
\Vertex(225,30){2}
\Text(225,5)[]{(d)}
\end{picture} 

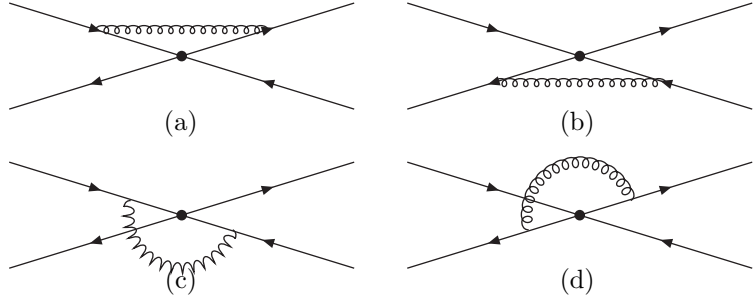
{Fig.2~~Feynman diagrams contributing to $Im{\cal M}(^{2S+1}L_J)_
{NRQCD}$ through the color-octet operator ${\cal O}_8(^{2S+1}(L-1)_{J^{\prime}}
)$.}

\end{center}

In the effective NRQCD theory, we can recalculate 
these annihilation amplitudes by considering the contributions of 4-fermion 
operators. 
The tree level contributions come from the diagram in Fig.1, where $Q\bar{Q}$ 
annihilate by color-singlet operator. The essential point is that some 
color-octet operators also contribute to the annihilation amplitudes through 
one-loop diagrams in Fig.2, which represent the transition between color-singlet 
component and color-octet component by emitting soft gluons. These color-octet operators
are ${\cal O}_8(^3S_1)$, ${\cal O}_8(^1S_0)$, ${\cal O}_8(^3P_0)$, and
${\cal O}_8(^3P_2)$.
By an explicit calculation we get
\begin{equation}\label{3pn}
Im{\cal M}(^3P_J)_{NRQCD}=\frac{Imf_1(^3P_J)}{m^6}+Imf_8(^3S_1)\frac{4C_F\alpha_s}
{3m^6N_c\pi}(-\frac{1}{2\epsilon_{IR}}+\frac{1}{2\epsilon_{UV}}),
\end{equation} 
\begin{equation}
Im{\cal M}(^1P_1)_{NRQCD}=\frac{Imf_1(^1P_1)}{m^6}+Imf_8(^1S_0)\frac{4C_F\alpha_s}
{3m^6N_c\pi}(-\frac{1}{2\epsilon_{IR}}+\frac{1}{2\epsilon_{UV}}),
\end{equation} 
\begin{equation}
Im{\cal M}(^3D_1)_{NRQCD}=\frac{Imf_1(^3D_1)}{m^6}+[Imf_8(^3P_0)\frac{20C_F\alpha_s}
{27m^6N_c\pi}+Imf_8(^3P_2)\frac{C_F\alpha_s}{27m^6N_c\pi}]
(-\frac{1}{2\epsilon_{IR}}+\frac{1}{2\epsilon_{UV}}),
\end{equation} 
\begin{equation}
Im{\cal M}(^3D_2)_{NRQCD}=\frac{Imf_1(^3D_2)}{m^6}+Imf_8(^3P_2)\frac{C_F\alpha_s}
{3m^6N_c\pi}(-\frac{1}{2\epsilon_{IR}}+\frac{1}{2\epsilon_{UV}}),
\end{equation} 
\begin{equation}\label{3dn}
Im{\cal M}(^3D_3)_{NRQCD}=\frac{Imf_1(^3D_3)}{m^6}+Imf_8(^3P_2)\frac{4C_F\alpha_s}
{3m^6N_c\pi}(-\frac{1}{2\epsilon_{IR}}+\frac{1}{2\epsilon_{UV}}),
\end{equation}
where both infrared (IR) and ultraviolet (UV) divergences appear. The UV 
divergences can be removed after the renormalization of color octet operators. 
IR divergences are due to the nonperturbative effect for the transition of 
$Q\bar{Q}$ by emitting soft gluons. We will see in the following that these IR diverdences 
just equal to those that appear in the annihilation amplitude calculated in the full 
theory. At leading order the 
coefficients of color-octet operators can be obtained by simply substituting
the color factors for the corresponding color-singlet ones in \cite{BBL}. They read
\begin{equation}\label{c3s}
(Imf_8(^3S_1))_0=\frac{n_f\pi\alpha_s^2}{6}, 
\end{equation}
\begin{equation}
(Imf_8(^1S_0))_0=\frac{\pi(N_c^2-4)\alpha_s^2}{4N_c},
\end{equation}
\begin{equation}
(Imf_8(^3P_0))_0=\frac{3\pi(N_c^2-4)\alpha_s^2}{4N_c},
\end{equation}
\begin{equation}\label{c3p2}
(Imf_8(^3P_2))_0=\frac{\pi(N_c^2-4)\alpha_s^2}{5N_c}.
\end{equation}
Substituting the leading order coefficients (\ref{c3s}--\ref{c3p2}) into 
(\ref{3pn}--\ref{3dn}) and comparing them with the results in the full theory 
(\ref{3pf})--(\ref{3d3f}), we find that the infrared divergences are cancelled
and finite coefficients for color-singlet operators of P-wave and D-wave decays
are obtained. It is interesting to note that IR divergences in the annihilation
amplitudes of color-singlet P-wave $Q\bar{Q}$ can be factored into the matrix
elements of color-octet S-wave operators, and in the D-wave cases IR 
divergences can be factored into the matrix elements of color-octet P-wave
operators, and at order of $\alpha_s^3$ the color-octet S-wave operators do not contribute
to the IR divergence parts of annihilation of D-wave $Q\bar{Q}$ pair.

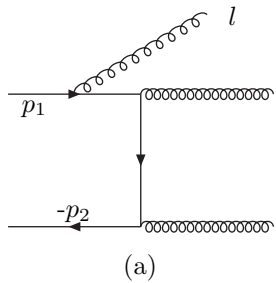
\begin{center}\begin{picture}(120,120)(0,0)
\ArrowLine(10,80)(60,80)\ArrowLine(60,80)(60,30)\ArrowLine(60,30)(10,30)
\Gluon(35,80)(85,110){2}{10}
\Gluon(60,80)(110,80){2}{15}\Gluon(60,30)(110,30){2}{15}
\Text(60,15)[]{(a)}
\Text(95,110)[]{$l$}\Text(20,75)[]{$p_1$}\Text(35,35)[]{-$p_2$}
\end{picture}
\end{center}

\begin{center}\begin{picture}(120,120)(0,0)
\ArrowLine(10,100)(60,100)\ArrowLine(60,100)(60,50)\ArrowLine(60,50)(10,50)
\Gluon(35,50)(85,20){2}{10}
\Gluon(60,100)(110,100){2}{15}\Gluon(60,50)(110,50){2}{15}
\Text(60,10)[]{(b)}
\Text(95,20)[]{$l$}\Text(35,95)[]{$p_1$}\Text(20,55)[]{$-p_2$}
\end{picture}
\end{center}

\begin{center}\begin{picture}(120,120)(0,0)
\ArrowLine(10,80)(60,80)\ArrowLine(60,80)(60,30)\ArrowLine(60,30)(10,30)
\Gluon(60,55)(110,55){2}{10}
\Gluon(60,80)(110,80){2}{15}\Gluon(60,30)(110,30){2}{15}
\Text(60,15)[]{(c)}
\Text(115,55)[]{$l$}\Text(35,65)[]{$p_1$}\Text(35,35)[]{-$p_2$}
\end{picture}

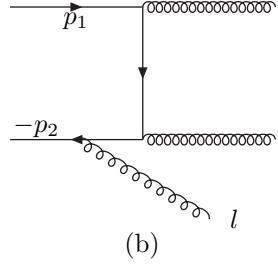
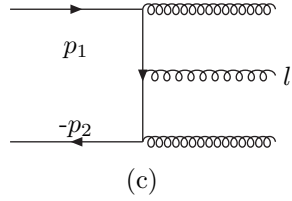
{Fig.3~~Examples of real-gluon emission in quarkonium decay at order
$\alpha_s^3$. Here $l$ represents the 4-momentum of a soft gluon.}
\end{center}

We can extend this conclusion
to a more general case that at leading nontrivial order in $v$ and at order
$\alpha_s^3$, the IR divergences appearing in the color-singlet 
$Q\bar{Q}(^{2S+1}L_J)$ annihilation amplitude can
just be cancelled by the contribution of color-octet operators 
${\cal O}(^{2S+1}(L-1)_{J^{\prime}})$,
${\cal O}(^{2S+1}(L-3)_{J^{\prime\prime}})$, $\cdots$.

Here we will give a detailed analysis for the cancellation 
mechanism. We know that at $\alpha_s^3$ the divergences come from
only two sources, $Q\bar{Q}\rightarrow ggg$ and $Q\bar{Q}\rightarrow
q_l\bar{q_l}g$. We give a specific example of decays into 3 gluons (Fig. 3) 
and the discussion applies also to decays into a
light quark-antiquark pair and a gluon. In the center of momentum frame, 
we take the  quarks to be on-shell with momentum $p_1$
and $p_2$ for $Q$ and $\bar{Q}$ respectively. By energy conservation there must
be at least two gluons with momenta of order $m$, so the diagram in Fig. 3(c)
contains no infrared divergence. For the diagrams in Fig. 3(a) and (b), the
emission vertex for the soft gluon with momentum $l$ and one adjacent 
heavy-quark propagator and one Dirac spinor can be approximated as follows:
\begin{eqnarray}\label{div1}\nonumber
\frac{\rlap/{p_1}-\rlap/{l}+m}{(p_1-l)^2-m^2}&\rlap/{\epsilon}&u(\vec{p})
\approx\frac{2p_1\cdot\epsilon}{-2p_1\cdot l}u(\vec{p})\\
&=&\frac{m\epsilon_{0}-\vec{p}\cdot\vec{\epsilon}}{-ml_0}(1+
\frac{\vec{p}\cdot\vec{l}}{ml_0}+\frac{(\vec{p}\cdot\vec{l})^2}{(ml_0)^2}
+\frac{(\vec{p}\cdot\vec{l})^3}{(ml_0)^3}+\cdots)u(\vec{p}),
\end{eqnarray}
\begin{eqnarray}\label{div2}\nonumber
\bar{v}(-\vec{p})&\rlap/{\epsilon}&\frac{\rlap/{l}-\rlap/{p_2}+m}{(l-p_2)^2-m^2}
\approx\frac{-2p_2\cdot\epsilon}{-2p_2\cdot l}\bar{v}(-\vec{p})\\
&=&\frac{-m\epsilon_{0}-\vec{p}\cdot\vec{\epsilon} }{-ml_0}(1-
\frac{\vec{p}\cdot\vec{l}}{ml_0}+\frac{(\vec{p}\cdot\vec{l})^2}{(ml_0)^2}
-\frac{(\vec{p}\cdot\vec{l})^3}{(ml_0)^3}+\cdots)\bar{v}(-\vec{p}),
\end{eqnarray} 
At leading nontrivial order in $v^2$ we must retain $L$ factors of $\vec{p}$ in the 
annihilation amplitudes of the $Q\bar{Q}$ pair whose orbital angular momentum
is $L$.  
The factor of $\vec{p}$ can come from one of two sources: the purely 
short-distance (infrared-safe) part of the diagram, or the potentially
infrared-divergent part (\ref{div1}) and (\ref{div2}), which consists of the soft gluon 
and the heavy-quark
propagator to which it attaches. If even factors of $\vec{p}$ come
from the infrared-divergent part of the diagram i.e., from (\ref{div1}) and
(\ref{div2}), it is obvious that the divergences cancel between the diagrams
of Fig.3(a) and Fig.3(b). But if odd factors of $\vec{p}$ from the divergence
part, the divergences come from the two diagrams add together, rather than
cancellation each other. 

In the decay rate, we must integrate the infrared emission factors from the square of the sum of the amplitudes over the phase space of gluon. We find results
\begin{eqnarray}\label{l1}\nonumber
I_1&=&4\sum_{polar}\int\frac{d^3l}{(2\pi)^32l_0}
\frac{(\epsilon_0\vec{p}\cdot\vec{l}-l_0\vec{p}\cdot\vec{\epsilon})}{ml_0^2}
\frac{(\epsilon_0\vec{p}^{\prime}\cdot\vec{l}-l_0\vec{p}^{\prime}
\cdot\vec{\epsilon})}{ml_0^2}\\
&=&\frac{2}{3m^2\pi^2}\vec{p}\cdot\vec{p}^{\prime}\int\frac{dl_0}{l_0},\\
\label{l3}
I_3&=&4\int\frac{d^3l}{(2\pi)^32l_0}\frac{(\vec{p}\cdot\hat{l})^2(\vec{p}^{\prime}
\cdot\hat{l})^2(\vec{p}\cdot\vec{p}^{\prime}-\vec{p}\cdot\hat{l}
\vec{p}^{\prime}\cdot\hat{l})}{m^6l_0^2}
\end{eqnarray}	
for the case of including one or three factors of $\vec{p}$ in it respectively.
Apart from the infrared emission factors, other propagator and vertex factors in Fig.3(a) and Fig.3(b) contribute to the short-distance part of the decay rate, which is equivalent to the annihilation of $Q\bar{Q}$ through 
color-octet operators ${\cal O}_8(^{2S+1}
(L-1)_{J^{\prime}})$ and ${\cal O}_8(^{2S+1}(L-3)_{J^{\prime\prime}})$ in NRQCD.

  In NRQCD the IR divergence parts of the correspondingly decay rates come from  one-loop diagrams in Fig.2 which include color-octet operators ${\cal O}_8(^{2S+1}
(L-1)_{J^{\prime}})$, ${\cal O}_8(^{2S+1}(L-3)_{J^{\prime\prime}})$, $\cdots$.
We can extract the IR divergence term through the integration
\begin{equation}
I=4i\int\frac{d^4q}{(2\pi)^4}\frac{\vec{p}\cdot\vec{p}^{\prime}
-\vec{p}\cdot\hat{q}\vec{p}^{\prime}\cdot\hat{q}}{m^2(q^2+i\varepsilon)}
\frac{1}{E-q_0-\frac{(\vec{p}-\vec{q})^2}{2m}+i\varepsilon}
\frac{1}{E-q_0-\frac{(\vec{p}^{\prime}-\vec{q})^2}{2m}+i\varepsilon}
\end{equation}
If one or three factors of $\vec{p}$ and $\vec{p}^{\prime}$ are retained, we
obtain
\begin{eqnarray}
I_1&=&\frac{2}{3m^2\pi^2}\vec{p}\cdot\vec{p}^{\prime}\int\frac{d|\vec{q}|}
{|\vec{q}|}\\
I_3&=&g^2\int\frac{d^3q}{(2\pi)^32|\vec{q}|}\frac{(\vec{p}\cdot\hat{q})^2(\vec{p}^{\prime}
\cdot\hat{q})^2(\vec{p}\cdot\vec{p}^{\prime}-\vec{p}\cdot\hat{q}
\vec{p}^{\prime}\cdot\hat{q})}{m^6|\vec{q}|^2}
\end{eqnarray}
It is obvious that they are the same as PQCD results (\ref{l1}) and
(\ref{l3}).
Therefore at order of $\alpha_s^3$ the source of divergence is
the transition of color-singlet $Q\bar{Q}(^{2S+1}L_J)$ to color-octet
$Q\bar{Q}(^{2S+1}(L-1)_{J^{\prime}})$, ${\cal O}_8(^{2S+1}(L-3)_{J^{\prime\prime}})$, $\cdots$ by emitting one, three $\cdots$ soft gluon, and these
divergences can be factored into the matrix elements of corresponding 
color-octet operators.  

In order to get a finite result at order $\alpha_s^3$ for the decay rates 
of high angular-momentum quarkonium states,
we must consider the contribution both from color-singlet and color-octet 
components. The complete formula for decay rate of P-wave states has been given
in \cite{huang}. In the following we will give the decay widths of D-wave 
triplet only to leading order in $\alpha_s$, and they can be written as
\begin{eqnarray}\nonumber
\Gamma(1^{--}\rightarrow LH)&=&2(Imf_8(^3P_0))_0\frac{<1^{--}|{\cal O}_8
(^3P_0)|1^{--}>}{m^4}+2(Imf_8(^3P_2))_0\frac{<1^{--}|{\cal O}_8
(^3P_2)|1^{--}>}{m^4}\\\nonumber
&=&\frac{152\pi(N^2_c-4)\alpha_s^2}{5N_c}H_P^8,
\\\nonumber
\Gamma(2^{--}\rightarrow LH)&=&2(Imf_8(^3P_2))_0\frac{<2^{--}|{\cal O}_8
(^3P_2)|2^{--}>}{m^4}\\
&=&\frac{18\pi(N^2_c-4)\alpha_s^2}{5N_c}H_P^8,
\\\nonumber
\Gamma(3^{--}\rightarrow LH)&=&2(Imf_8(^3P_2))_0\frac{<3^{--}|{\cal O}_8
(^3P_2)|3^{--}>}{m^4}\\
&=&\frac{72\pi(N^2_c-4)\alpha_s^2}{5N_c}H_P^8,
\end{eqnarray}
where we have neglected the contributions of the matrix elements of 
color-octet S-wave operators in NRQCD. We don't expect the S-wave component 
to play a major role in the decay od D-wave quarkonium in leading order in 
$\alpha_s$ according to the reason in \cite{cho}. Here we have used the 
heavy-quark spin-symmetry, which implies that, to leading order in $v^2$, 
the matrix element of ${\cal O}_8(^3P_{J^{\prime}})$ in the D-wave triplet
can be expressed as one nonperturbative parameter
$$
H_P^8=\frac{<1^{--}|{\cal O}_8(^3P_2)|1^{--}>}{m^4},
$$
which can be determined from lattice calculation or extracted by comparing with 
experimental data. Here we also have used the relation that rate of the matrix 
element $\frac{<J^{--}|{\cal O}_8(^3P_{J^{\prime}})|J^{--}>}{m^4}$ is $1:9:36$ 
for $J^{\prime}=2$ and $J=1,2,3$; and $1:20$ for $J=1$ and $J^{\prime}=2,0$.
This relation can be derived by considering the rate of E1 transition from 
color-singlet $^3D_J$ to color-octet $^3P_{J^{\prime}}$, which is proporational
to $(2J^{\prime}+1)\left\{\begin{array}{lll}1&1&2\\ 1&J&J^{\prime} \end{array}
\right\}^2$, where $\left\{\begin{array}{lll}1&1&2\\ 1&J&J^{\prime} \end{array}
\right\}$ represents the corresponding 6j coefficient.

It is interesting to indicate that the ratio for the decay rates of 
D-wave triplet is $R=76/9:1:4$, which is consistent with previous calculations
\cite{dwave}. The previously calculated $R$ come from the divergence parts of 
decay rate of color-singlet D-wave component, which are now factored into the
color-octet P-wave matrix elements, and to leading order in $\alpha_s$, we
only consider the contributions of color-octet P-wave components. Therefore
we derive the same $R$ as the previous one. This also happens in the 
P-wave case \cite{pwave}.
     
In the above discussions we have only studied the quarkonium decays at leading
nontrivial order in $v^2$ and to next-to-leading order in $\alpha_s$.
In principle, it can be applied to any given order in $v^2$ and $\alpha_s$, because it is 
based on a solid theoretical formalism. But in high order cases the problem
is very complicated and more operators have to be considered.
\vfill\eject

\end{document}